\documentclass[preprint,floats,aps,epsfig,nofootinbib,amssymb]{revtex4}
\pdfoutput=1
\usepackage{mathrsfs}
\usepackage{slashed}
\usepackage{graphicx,color}
\usepackage{epsfig}
\usepackage{subfigure}
\usepackage{epsfig}
\usepackage{dcolumn}
\usepackage{bm}

\begin{document}

\title{Testing lepton flavor universality in terms of data of BES III and charm-tau factory}

\author{\bf  Bin Wang$^1$, Ming-Gang Zhao$^1$, Ke-Sheng Sun$^2$  and Xue-Qian Li$^1$}

\affiliation{1. School of Physics, Nankai University, Tianjin, 300071, China,\\
2. School of Physics, Hebei University, Baoding, Hebei Province, 071002, China. }

\begin{abstract}
The recent measurements on $R_K$ and $R_{\pi}$ imply that there exists a possible violation of the leptonic flavor universality which is one of the cornerstones of the standard model. It is suggested that a mixing between sterile and active neutrinos might induce such a violation. In this work we consider the scenarios with one or two sterile neutrinos to explicitly realize the data while the constraints from the available experiments have been taken into account. Moreover, as indicated in literature, the deviation of the real PMNS matrix from the symmetric patterns may be due to a $\mu-\tau$ asymmetry, therefore the measurements on $R_{D(D_s)e\mu}=\Gamma(D(D_s)\rightarrow e^+\nu_e)/\Gamma(D(D_s)\rightarrow \mu^+\nu_\mu)$ and
$R_{D(D_s)\mu\tau}=\Gamma(D(D_s)\rightarrow \mu^+\nu_\mu)/\Gamma(D(D_s)\rightarrow \mu^+\tau_\tau)$ (and for some other heavy mesons $B^{\pm}$ and $B_c$ etc.) may shed more light on physics responsible for the violation of the leptonic flavor universality. The data of BES III are available to test the universality and that of the future charm-tau factory will provide more accurate information towards the aspect,  in this work, we will discuss $R_{D(D_s)e\mu}$ and $R_{D(D_s)\mu\tau}$  in all details and also briefly consider the cases for $B^{\pm}$ and $B_c$.



\end{abstract}

\draft

\maketitle

\section{Introduction}
The property that couplings of leptons to gauge bosons are independent of lepton flavors is named as lepton flavor universality (LFU), which is embedded in the Standard Model (SM) and any violation of LFU may be induced by new physics beyond the SM. Taking leptonic decays of $W$ boson $W\rightarrow l\nu_l$ ($l=e$, $\mu$, $\tau$) as an instance, ratios of the branching fractions measured at LEP II \cite{LEPII} are
\begin{eqnarray}
&&B(W\rightarrow \mu \overline{\nu}_\mu) / B(W\rightarrow e \overline{\nu}_e)=0.997\pm 0.021, \\
&&B(W\rightarrow \tau \overline{\nu}_\tau) / B(W\rightarrow e \overline{\nu}_e)=1.058\pm 0.029, \\
&&B(W\rightarrow \tau \overline{\nu}_\tau) / B(W\rightarrow \mu \overline{\nu}_\mu)=1.061\pm 0.028.
\end{eqnarray}
which demonstrate the lepton universality at the level of $2.9\%$. Besides, leptonic decays of mesons also provide a possibility to test LFU where the uncertainties in the hadronic sector are cancelled. In literature the ratios
are suggested to be measured
\begin{eqnarray}
R_{P\alpha\beta} \equiv {{\Gamma(P^+\rightarrow \alpha^+ \nu_\alpha)}\over{\Gamma(P^+\rightarrow\beta^+ \nu_\beta)}},
\end{eqnarray}
where $P=\pi$, $K$, $D$, $D_s$, $B$, $B_c$ and $\alpha$, $\beta=e$, $\mu$, $\tau$. In order to clearly demonstrate deviation of the measured value $R_{P\alpha\beta}^{\text{exp}}$ from the SM predictions $R_{P\alpha\beta}^{\text{SM}}$, a parameter $\Delta r_{P\alpha\beta}$ is defined as
\begin{eqnarray}
\Delta r_{P\alpha\beta} \equiv {{R_{P\alpha\beta}^{\text{exp}}}\over{R_{P\alpha\beta}^{\text{SM}}}} -1.
\end{eqnarray}

To accommodate non-zero $\Delta r_{P\alpha\beta}$, one may invoke two different mechanisms \cite{PiKsterile}:
\begin{itemize}
\item Introducing new Lorentz structure in the four-fermion interaction;
\item Modifying the $Wl\nu_l$ vertex by corrections to lepton mixing.
\end{itemize}
For the first category, SM may be extended to new physics beyond standard model (BSM) which includes charged Higgs, for example in the supersymmetry (SYSY) \cite{chargeHiggs} or two-Higgs-doublet models \cite{2HDM}. In the SM, the violation of LFU are estimated as $|\Delta r_{P\alpha\beta}^{\text{SM}}|=\mathcal{O}({{\alpha}\over{4\pi}}\times {{m_{\alpha(\beta)}^2}\over{m_W^2}})$, which is too small to be accounted, while for the first category, the correction to the $W\alpha\nu_\alpha$ vertex emerges at loop level where new physics BSM particles exist  is of order $\mathcal{O}({{\alpha}\over{4\pi}}\times {{m_W^2}\over{\Lambda_{\text{NP}}^2}})$, which is greatly suppressed by the new physics scale $\Lambda_{\text{NP}}$ \cite{susy1loop}.


For the second category, the coupling $W\alpha\nu_\alpha$ is modified by breaking the unitarity of the $3\times 3$ lepton mixing matrix. In the type-I seesaw \cite{typeIseesaw}, sterile neutrinos are introduced and the $3\times 3$ mixing matrix could be non-unitary to the level of $\mathcal{O}(M^2_D/M^2_R)$ \cite{XZhBook}\cite{InverseSeesaw}, where $m_D$ and $m_R$ are the Dirac and Majorana mass matrices, respectively.

Introducing sterile neutrinos can provide natural interpretations of some anomalies observed in recent experiments. For instance, the LSND \cite{LSNDanomaly}, the MiniBooNE \cite{MiniBooNE}, the reactor \cite{ReactorAnomaly} anomalies and as well as the
gallium anomaly of the GALLEX \cite{GALLEX} and SAGE \cite{SAGE} experiments. Furthermore, a careful analysis of the cosmological data show that the effective number of neutrino species is larger than 3  \cite{GongBoZhao}, which might also hint existence of additional one or two neutrino species besides the three active ones. Therefore a study on violation of LFU would be interesting because it may reveal possible new physics mechanisms beyond the SM at the lepton sector.

As indicated in literature, the deviation of the real PMNS matrix from the symmetric patterns may be due to a $\mu-\tau$ asymmetry. Since the kinematic constraints, $K$ and $\pi$ cannot decay into $\tau\nu$,
the measurements on $R_{D(D_s)e\mu}=\Gamma(D(D_s)\rightarrow e^+\nu_e)/\Gamma(D(D_s)\rightarrow \mu^+\nu_\mu)$ and
$R_{D(D_s)\mu\tau}=\Gamma(D(D_s)\rightarrow \mu^+\nu_\mu)/\Gamma(D(D_s)\rightarrow \mu^+\tau_\tau)$ (and some other heavy mesons $B^{\pm}$ and $B_c$ etc.) may shed more light on the physics responsible for the violation of the leptonic flavor universality. The data of BES III are available to test the universality and measurements at the future charm-tau factory will provide more accurate information towards the aspect. Indeed, the experimental results of NA62 Collaboration \cite{Na62} determine $\Delta r_{Ke\mu}=(4\pm 4)\times 10^{-3}$ and $\Delta r_{\pi e\mu}=(-4\pm 3)\times 10^{-3}$ \cite{PiKsterile} whereas the BES data tell us $\Delta r_{D_s\mu\tau}=0.256^{+0.430}_{-0.317}$ which is rather large.

The presence of sterile neutrinos changes the leptonic decay widths of pseudoscalar mesons and then  $\Delta r_{P\alpha\beta}$ may be enhanced or suppressed. In this work we start with this motivation and derive the analytical formulas of $\Delta_{P\alpha\beta}$ for 3+1 and 3+2 scenarios in the next section. Especially in this work, we will discuss $R_{D(D_s)e\mu}$ and $R_{D(D_s)\mu\tau}$  in all details and also briefly consider the cases for $B^{\pm}$ and $B_c$. Numerical analyses are made and possible experimental measurements on $\Delta r_{P\alpha\beta}$ at BES and future charm-tau factory are discussed. Then we present some discussions in the last section.

\section{$R_{P\alpha\beta}$ within SM}
In the SM, the leptonic decay widths of pseudo-scalar mesons $P\rightarrow \alpha\nu_\alpha$ ($\alpha=e$, $\mu$ or $\tau$ and $P=\pi$, $K$, $D$, $D_s$, $B$ or $B_c$) are given by
\begin{eqnarray}
\Gamma_{P\rightarrow \alpha\nu_\alpha}^{\text{SM}} = {{G_F^2}\over{8\pi}}|V_{qq^\prime}|^2 f_P^2 m_P m_\alpha^2 \left(1-{{m_\alpha^2}\over{m_P^2}}\right)^2, \label{decaySM}
\end{eqnarray}
where $G_F$ is the Fermi constant, $m_\alpha$ is the lepton mass, $V_{qq^\prime}$ the element of Cabibbo-Kobayashi-Maskawa (CKM) mixing matrix corresponding to the constituents in the meson $P$ whose mass is $m_P$ and decay constant $f_P$ and neutrino masses are neglected. As well known, these processes are helicity suppressed.

With the definition of $\Gamma_{P\rightarrow \alpha\nu_\alpha}^{\text{SM}}$ in Eq. (\ref{decaySM}), one has
\begin{eqnarray}
R_{P\alpha\beta}^{\text{SM}}\equiv \left({m_\alpha\over m_\beta}\right)^2 \left({{m_P^2-m_\alpha^2}\over{m_P^2-m_\beta^2}}\right)^2. \label{RatioSM}
\end{eqnarray}
If the QED corrections are considered, there will be a factor $(1+\delta_{QED})$ multiplying to the left-handed side of Eq. (\ref{RatioSM}).
An approximate estimate of electrodynamical correction to $P\rightarrow \alpha\nu_\alpha$ is about ${1\over{137}}\times {1\over{2\pi}}\sim 10^{-3}$ suppression. For the leptonic kaon decay, due to the inner bremsstrahlung $K_{l2\gamma}$ process which is included by definition into $R_K$, the calculation indicates $\delta_{QED}=(-3.78\pm0.04)\%$ \cite{KaonQED}.

As suggested, existence of sterile neutrinos  changes the picture, so in the next section, we will study how they contribute to $P\rightarrow \alpha\nu_\alpha$.

\section{Lepton mixing with presence of sterile neutrinos}

Now let us give a general description of the involvement of sterile neutrinos and take 3+1 as an illustration. In the SM, the sterile neutrinos do not directly couple to $W$-boson, thus without a mixing between sterile neutrinos with active ones, the weak interaction can be written as
\begin{equation}
\overline{(e, \mu, \tau)}\gamma^{\mu}(1-\gamma_5)(U_{\text{PMNS}},0)
\left(
\begin{array}{c}
\nu_1 \\ \nu_2 \\ \nu_3 \\ \nu_{s}
\end{array}
\right)W_{\mu},
\end{equation}
where $\nu_i (i=1,2,3)$ are the active neutrinos and $\nu_{s}$ is a sterile neutrino (or several), $U_{\text{PMNS}}$ is the regular PMNS lepton mixing matrix and here $0$ denotes a 3$\times 1$ matrix, thus $(U_{\text{PMNS}},0)$ is a $3\times 4$ matrix and in this scenario, unitarity is no longer existing. When the mixing between $\nu_{s}$ and $\nu_i$ is introduced, the matrix becomes $(U_{\text{PMNS}}\cdot\cos\epsilon, \sin\epsilon)$ and the coupling vertex is
\begin{equation}
\overline{( e, \mu, \tau)}\gamma^{\mu} (1-\gamma_5)(U_{\text{PMNS}}\cdot \cos\epsilon, \sin\epsilon)
\left(
\begin{array}{c}
\nu_1 \\ \nu_2 \\ \nu_3 \\ \nu_{s}
\end{array}
\right)W_{\mu},
\end{equation}
then the sterile neutrino participates in the weak interaction\footnote{After the replacement, $\nu_1,\nu_2,\nu_3,\nu_s$ are slightly changed and constitute the real mass eigenstate.}. Meanwhile this mixing also induces a flavor-changing neutral current (FCNC) as
\begin{eqnarray}
\overline{(\nu_1, \nu_2,  \nu_3, \nu_{s})}\gamma^{\mu}(1-\gamma_5)(1,0) \left(
\begin{array}{c}
\nu_1 \\ \nu_2 \\ \nu_3 \\ \nu_{s}
\end{array}
\right)Z_{\mu}
\rightarrow \overline{(\nu_1, \nu_2, \nu_3, \nu_{s1})}\gamma^{\mu}(1-\gamma_5)V'
\left(
\begin{array}{c}
\nu_1 \\ \nu_2 \\ \nu_3 \\ \nu_{s}
\end{array}
\right)Z_{\mu},
\end{eqnarray}
where $V'$ is a $4\times 4$ matrix and resulted in from a matrix product of
$(4\times 3)\otimes (3\times 4)$ i.e. $(U_{\text{PMNS}}\cdot\cos\epsilon, \sin\epsilon)^{\dagger}(U_{\text{PMNS}}\cdot\cos\epsilon, \sin\epsilon)$. This non-unitary mixing matrix and sizable sterile neutrino masses can also change the generation number which is accurately measured by the LEP experiment. Therefore the mixing is rigorously constrained by the LEP data.

Once sterile neutrinos are introduced, they do mix with active ones, the decay width obtained above (\ref{decaySM}) will be modified into
\begin{eqnarray}
&&\Gamma_{P\rightarrow \alpha\nu_\alpha}^{\text{s}}={{G_F^2}\over{8\pi}} |V_{pp^\prime}|^2 f_P^2 m_P \times\\ \nonumber
&&\left[\sum_{i=1}^3|U_{\alpha i}|^2 m_\alpha^2(1-{m_\alpha^2\over{m_P^2}})^2+ \sum_{i=1}^{N-3} |U_{\alpha, i+3}|^2 (m_\alpha^2+m_{si}^2) \left(1-{{m_\alpha^2+m_{si}^2}\over{m_P^2}}\right) {{\lambda(m_P^2, m_\alpha^2, m_{si}^2)}\over{m_P^2}}\right],\label{decaySterile}
\end{eqnarray}
where $m_{si}$ is the sterile neutrino masses and $N$ is the total number of neutrino mass eigenstates, for instance, assuming two sterile neutrinos, i.e., the so-called $3+2$ scenario \cite{WhitePaper}, $N=3+2=5$. The $\lambda(m_P^2, m_\alpha^2, m_{si}^2)$ is defined as
\begin{eqnarray}
\lambda(m_P^2, m_\alpha^2, m_{si}^2)=
\sqrt{m_P^4+m_\alpha^4+m_{si}^4 - 2m_P^2m_\alpha^2-2m_P^2m_{si}^2-m_\alpha^2m_{si}^2}.
\end{eqnarray}

Now we can obtain the general expression of $\Delta_{P\alpha\beta}$
\begin{eqnarray}
\Delta r_{P\alpha\beta} = {{R_{P\alpha\beta}^{\text{s}}}\over{R_{P\alpha\beta}^{\text{SM}}}} - 1, \label{Delatar}
\end{eqnarray}
where $R_{P\alpha\beta}^{\text{SM}}$ is obtained in Eq. (\ref{RatioSM}) and
\begin{eqnarray}
&&R_{P\alpha\beta}^{\text{s}}= \nonumber\\
&&{{\sum_{i=1}^3|U_{\alpha i}|^2m_\alpha^2(m_P^2-m_\alpha^2)^2+ \sum_{i=1}^{N-3} |U_{\alpha, i+3}|^2 (m_\alpha^2+m_{si}^2) \left(m_P^2-m_\alpha^2-m_{si}^2\right) \lambda(m_P^2, m_\alpha^2, m_{si}^2)}\over{\sum_{j=1}^3|U_{\beta j}|^2 m_\beta^2(m_P^2-m_\beta^2)^2+ \sum_{j=1}^{N-3} |U_{\beta, j+3}|^2 (m_\beta^2+m_{sj}^2) \left(m_P^2-m_\beta^2-m_{sj}^2\right) \lambda(m_P^2, m_\beta^2, m_{sj}^2)}}. \nonumber \\ \label{RatioSterile}
\end{eqnarray}

\subsection{3+1 scenario}
In order to get the $3\times 4$  mixing matrix describing the sterile-active neutrino mixing in charged-current and neutral-current interactions, an easy means is to construct a $4\times 4$ matrix in a regular way and then remove away the last row of the  $4\times 4$ matrix to get the required $3\times 4$ matrix. The $3\times 5$ matrix for the 3+2 scenario will be obtained in a similar way.

First we deal with the case by adding one sterile neutrino into the game, i.e., the $3+1$ scenario, in which the lepton mixing matrix can be parameterized with 6 mixing angles and 3 phases in a form
\begin{eqnarray}
U_{3+1}=R_{34}(\theta_{34}, \delta_{34}) R_{24}(\theta_{24}, 0) R_{14}(\theta_{14}, \delta_{14}) R_{23}(\theta_{23}, 0) R_{13}(\theta_{13}, \delta_{13}) R_{12}(\theta_{12}, 0), \label{3plus1}
\end{eqnarray}
where $R_{ij}$ is a $4\times 4$ matrix describing the rotation in the $i-j$ plane and here presents $R_{34}$ as an instance,
\begin{eqnarray}
R_{34}(\theta_{34}, \delta_{34})=\left(
\begin{array}{cccc}
1 & 0 & 0 & 0 \\
0 & 1 & 0 & 0 \\
0 & 0 & \cos\theta_{34} & \sin\theta_{34}e^{-i\delta_{34}} \\
0 & 0 & -\sin\theta_{34}e^{i\delta_{34}} & \cos\theta_{34} \end{array}
\right). \label{R34}
\end{eqnarray}
It is obvious that the multiplication $R_{23}R_{13}R_{12}\equiv U^0$ is the usual Pontecorvo \cite{PMNS1}-Maki-Nakawaga-Sakata \cite{PMNS2} (PMNS) matrix modified by adding a trivial fourth column and row.

In order not to be contradict with the experimental measurements of available three-neutrino oscillations and the data of LEP, the mixing between sterile and active neutrinos must be very small. And here an assumption is made that the added sterile neutrino does not distinguish between the three active ones, i.e., the mixing angles $\theta_{34}=\theta_{24}=\theta_{14}\equiv \epsilon_1$. In Eq. (\ref{RatioSterile}) only squared mixing elements $|U_{\alpha i}|^2$ appear in $\Delta r_{P\alpha\beta}$ while neglecting higher order powers of $\epsilon_1$, one obtains
\begin{eqnarray}
&&|U_{\alpha i}|^2=|U_{\alpha i}^0|^2\cos^2\epsilon_1, (\alpha=e,\mu,\tau; i=1,2,3); \nonumber \\
&&|U_{e4}|^2=\sin^2\epsilon_1;|U_{\mu4}|^2=\sin^2\epsilon_1\cos^2\epsilon_1;
|U_{\tau4}|^2=\sin^2\epsilon_1\cos^4\epsilon_1,
\end{eqnarray}
where we ignore the subscript "3+1" in $U_{3+1}$ and the $U_{\alpha i}^0$ refers to the elements of the unitary $U^0$ (3+0)  with $\sum_{i=1}^3 |U_{\alpha i}^0|^2=1$ for $\alpha=e, \mu, \tau$.

Then with these mixing matrix elements, one can obtain $R_{P\alpha\beta}^{\text{s}}$ in the 3+1 scenario
\begin{eqnarray}
R_{P\alpha\beta}^{\text{s}}=
{{\cos^2\epsilon_1 m_\alpha^2(m_P^2-m_\alpha^2)^2+ |U_{\alpha 4}|^2 (m_\alpha^2+m_{s1}^2) \left(m_P^2-m_\alpha^2-m_{s1}^2\right) \lambda(m_P^2, m_\alpha^2, m_{s1}^2)}\over{\cos^2\epsilon_1 m_\beta^2(m_P^2-m_\beta^2)^2+ |U_{\beta 4}|^2 (m_\beta^2+m_{s1}^2) \left(m_P^2-m_\beta^2-m_{s1}^2\right) \lambda(m_P^2, m_\beta^2, m_{s1}^2)}}, \nonumber \\ \label{Ratio3p1}
\end{eqnarray}
which is substituted into Eq. (\ref{Delatar}) to obtain $\Delta r_{P\alpha\beta}$ under the 3+1 scenario.

\subsection{3+2 scenario}
Turning to the 3+2 scenario in which two sterile neutrinos are introduced, the procedure is similar to that for the 3+1 case. The mixing angles between the second sterile neutrino and active neutrinos are denoted as $\epsilon_2$. Then we summarize the results below:
\begin{eqnarray}
&&
|U_{\alpha i}|^2\simeq|U_{ei}^0|^2\cos^2\epsilon_1\cos^2\epsilon_2, (\alpha=e,\mu,\tau; i=1,2,3);\nonumber \\
&&
|U_{e4}|^2=\sin^2\epsilon_1\cos^2\epsilon_2;
|U_{e5}|^2=\sin^2\epsilon_2;
\nonumber \\
&&
|U_{\mu4}|^2=\sin^2\epsilon_1\cos^2\epsilon_1\cos^2\epsilon_2;
|U_{\mu5}|^2=\sin^2\epsilon_2\cos^2\epsilon_2;
\nonumber \\
&&
|U_{\tau4}|^2\simeq\sin^2\epsilon_1\cos^4\epsilon_1\cos^2\epsilon_2;
|U_{\tau5}|^2=\sin^2\epsilon_2\cos^4\epsilon_2.
\end{eqnarray}
With these elements one can obtain $\Delta r_{P\alpha\beta}$ with the 3+2 scenario.

\subsection{Constraints set by the FCNC }

With the scenarios discussed above, the neutrino neutral current (NC) interaction will be modified. As discussed above,
the number of neutrinos is $2.984\pm 0.008$ by fits to LEP data. Assuming the small deviation from 3 to be caused by mixing between active and sterile neutrinos, this value will cast rigorous constraints on the mixing parameter $\epsilon_{1, 2}$. An estimate is: $\epsilon_1\leq \mathcal{O}(5\times 10^{-2})$ for the 3+1 scenario and $\epsilon_1^2+\epsilon_2^2\leq \mathcal{O}(10^{-3})$ for the 3+2 case. A preliminary result about sum $\sum_{i=1}^{3}|U_{ei}|=0.994\pm 0.005$ at $90\%$ confidence level \cite{VeiBound}, which signifies the non-unitarity of the $3\times 3$ active neutrino mixing matrix to be at level $\leq \mathcal{O}(10^{-2})$ \cite{Xing1percent}. The non-vanishing terms $U_{\alpha\beta}\equiv\sum_{i=4}^{N} U^*_{\alpha i} U_{i\beta}$ ($\alpha\neq\beta$) result in the tree level FCNC interaction, which can induce the low-energy lepton flavor violation (LFV) processes. And these LFV precesses are proportional to the value of the non-vanishing $\sum_{i=4}^{N} U^*_{\alpha i} U_{i\beta}$. The experimental bounds on these LFV interactions can be transformed to constraints on the new mixing angles $\epsilon_{1, 2}$. For instance, constraints to $|U_{\alpha \beta}|$ from several LFV processes Ref. \cite{PDG,UBound}:
\begin{eqnarray}
&& |U_{\mu e}|<3.05\times 10^{-6} \left(\mathcal{B}(\mu^-\rightarrow e^-e^+e^-)<1.0\times 10^{-12}\right), \nonumber \\
&& |U_{\tau e}|<1.37\times 10^{-3} \left(\mathcal{B}(\tau^-\rightarrow e^-e^+e^-)<3.6\times 10^{-8}\right), \nonumber \\
&& |U_{\tau\mu}|<1.295\times 10^{-3} \left(\mathcal{B}(\tau^-\rightarrow \mu^-\mu^+\mu^-)<3.2\times 10^{-8}\right) \nonumber .
\end{eqnarray}
Considering these bounds and the elements derived above, with reasonable approximations we can obtain constraints on $\epsilon_{1, 2}$ which are shown in Table \ref{TabBound}.
\begin{table}
\begin{center}
\begin{tabular}{|c||c|c|c|}
\hline
Scenario & $|U_{\mu e}|$ & $|U_{\tau e}|$ & $|U_{\tau \mu}|$ \\
\hline\hline
3+1 & $\epsilon_1<1.75\times 10^{-3}$ & $\epsilon_1<3.70\times 10^{-2}$ & $\epsilon_1<3.599\times 10^{-2}$ \\
\hline
3+2 & $\epsilon_1^2+\epsilon_2^2<3.05\times 10^{-6}$ & $\epsilon_1^2+\epsilon_2^2<1.37\times 10^{-3}$ & $\epsilon_1^2+\epsilon_2^2<1.295 \times 10^{-3}$ \\
\hline
\end{tabular}
\begin{quote}
\caption{Constraints to new added mixing parameters $\epsilon_{1, 2}$ from the experimental limits of LFV processes.}
\label{TabBound}
\end{quote}
\end{center}
\end{table}

\section{Numerical analyses}
In this section we numerically evaluate the total contributions from SM and new physics BSM to $\Delta r_{P\alpha\beta}$ in the context of 3+1 and 3+2 scenarios to account for the observed data. The analyses are about the leptonic decays of pseudo-scalar mesons  $P=\pi, K, D, D_s, B, B_c$. The evaluated branching ratios of the leptonic decays are listed in Table \ref{MesonParameter} and the  corresponding $R_{P\alpha\beta}$ and $\Delta r_{P\alpha\beta}$ are shown in Table \ref{TabRDr}.
\begin{table}
\begin{center}
\begin{tabular}{|c||c|c|c|c|}\hline
$P$ & mass (MeV) & $\mathcal{B}(P\rightarrow e\nu_e)$ & $\mathcal{B}(P\rightarrow \mu\nu_\mu)$ & $\mathcal{B}(P\rightarrow \tau\nu_\tau)$ \\
\hline\hline
$\pi$ & $139.57018\pm 0.00035$ & $(1.230\pm0.004)\times 10^{-4}$ & $(99.98770\pm 0.00004)\%$ & - \\
\hline
$K$ & $(493.677\pm 0.016)$ & $(1.581\pm 0.008)\times 10^{-5}$ & $(63.55\pm 0.11)\%$ & - \\
\hline
$D$ & $(1869.62\pm 0.15)$ & $<8.8\times 10^{-6}$ & $(3.82\pm 0.33)\times 10^{-4}$  & $<1.2\times 10^{-3}$ \\
 & & & $[(4.15_{-0.21}^{+0.22})\times 10^{-4}]$ &$[(1.10\pm 0.06)\times 10^{-3}]$  \\
\hline
$D_s$ & $1968.49\pm 0.32$ & $<1.2\times 10^{-4}$ & $(5.90\pm 0.33)\times 10^{-3}$ & $(5.43\pm 0.31)\%$ \\
 & & & $[(5.50_{-0.52}^{+0.55})\times 10^{-3}]$ & $[(5.36_{-0.50}^{+0.54})\times 10^{-2}]$ \\
\hline
$B$ & $5279.25\pm 0.17$ & $<9.8\times 10^{-7}$ & $<1.0\times 10^{-6}$ & $(1.65\pm 0.34)\times 10^{-4}$ \\
 & & & & $[(0.796_{-0.087}^{+0.088})\times 10^{-4}]$ \\
 & & $[[(1.1\pm 0.2)\times 10^{-11}]]$ &  $[[(4.5\pm 1.0)\times 10^{-7}]]$ & $[[(1.0\pm 0.2)\times 10^{-4}]]$ \\
\hline
$B_c$ & $6277\pm 6$ & - & - & - \\
\hline
\end{tabular}
\begin{quote}
\caption{The experimental values or bounds of masses and branching ratios for $P=\pi, K, D, D_s, B, B_c$ taken from PDG \cite{PDG}. The corresponding theoretical predictions are in (double) square brackets which are from Ref. (\cite{BEMuTau}) \cite{RPiK}. }
\label{MesonParameter}
\end{quote}
\end{center}
\end{table}

\begin{table}
\begin{center}
\begin{tabular}{|c||c|c|c|}\hline
$P$ & $R_{Pe\mu}^{\text{exp}}$ & $R_{Pe\mu}^{\text{SM}}$ & $\Delta r_{Pe\mu}$\\
\hline
$\pi$ & $(1.230\pm 0.004)\times 10^{-4}$ & $1.234\times 10^{-4}$ & $(-3.241\pm 3.241)\times 10^{-3}$\\
\hline
$K$ & $(2.488\pm 0.013)\times 10^{-5}$ & $(2.472\pm 0.001)\times 10^{-5}$ & $(6.472_{-5.664}^{+5.668})\times 10^{-3}$ \\
\hline \hline \hline
$P$ & $R_{P\mu\tau}^{\text{exp}}$ & $R_{P\mu\tau}^{\text{SM}}$ & $\Delta r_{P\mu\tau}$ \\
\hline
$D$ & $>0.291$ & $0.377_{-0.038}^{+0.043}$ & $>-0.308$\\
\hline
$D_s$ & $0.109_{-0.012}^{+0.013}$ & $(8.65_{-1.43}^{+1.68})\times 10^{-2}$ & $0.256_{-0.317}^{+0.430}$ \\
\hline
$B$ & $<7.6\times 10^{-3}$ & $(4.5_{-1.6}^{+2.4})\times 10^{-3}$ & $<1.6$ \\
\hline
$B_c$ & - & $(4.18_{-0.04}^{+0.03})\times 10^{-3}$ & - \\
\hline
\end{tabular}
\begin{quote}
\caption{The current experimental measurements and SM prediction of $R_{Pe\mu}$ \cite{RPiK}, $R_{P\mu\tau}$ and the corresponding $\Delta r_{P\alpha\beta}$. The SM predictions include
the uncertainties from electromagnetic corrections and as well as the uncertainties due to CKM mixing matrix elements and decay constants.}
\label{TabRDr}
\end{quote}
\end{center}
\end{table}

\subsection{3+1 scenario}
\begin{figure}[t]
\centering
\includegraphics[height=8cm, width=16cm, angle=0]{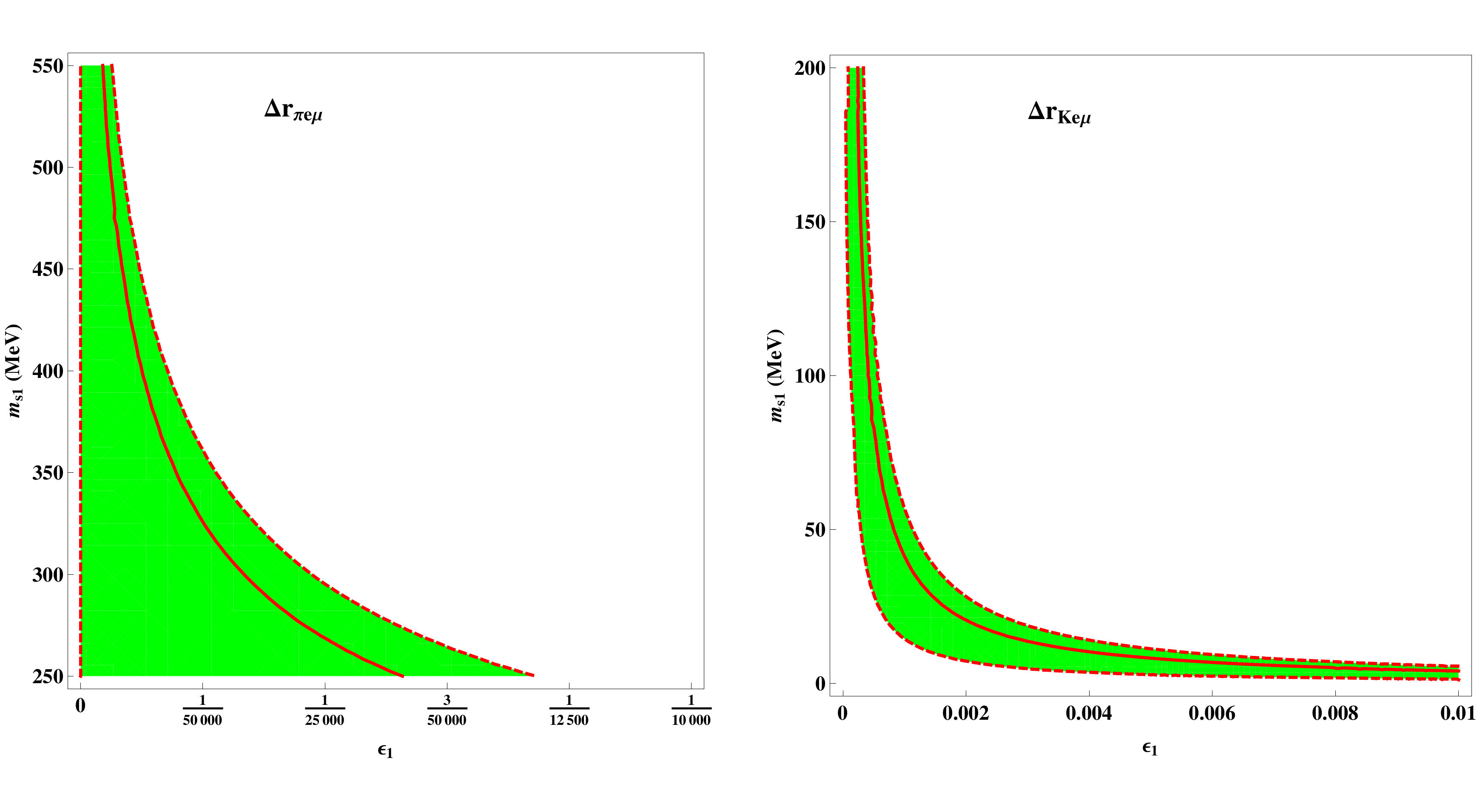}
\caption{The allowed parameter space of $\epsilon_1$ and $m_{s1}$ for $\Delta r_{\pi e\mu}$ and $\Delta r_{Ke\mu}$.} \label{FigDrPiKabc}
\end{figure}

In this subsection, we make a numerical analysis on the breaking of the lepton universality in different decay processes,
within the $3+1$ scenario.
$\Delta r_{\pi e\mu}$ and $\Delta r_{K e\mu}$ in the parameter space of mixing parameter $\epsilon_1$ and sterile neutrino mass $m_{s1}$ are shown in Fig. \ref{FigDrPiKabc}.
The central value of $\Delta r_{Pe\mu}$ is denoted by the solid line while its $1-\sigma$ region is enclosed by the dashed lines. In the left panel of Fig. \ref{FigDrPiKabc} for $\Delta r_{\pi e\mu}$, to accommodate the experimental results, the mass of the sterile neutrino has to be larger than 250 MeV,  thus the final state with $\nu_s$ is kinematically forbidden. For $\Delta r_{Ke\mu}$, the allowed parameter space of $\epsilon_1$ and $m_{s1}$ is larger, which covers $0\sim 400$ MeV and $0\sim 0.35$, respectively. Also, the region of $m_{s1}>m_K$, $\nu_s$ does not show up in the final state  and the region with relatively larger $\epsilon_1$ needs to be ruled out for its failure to be reconciled with the LEP data constraint.

\begin{figure}[t]
\centering
\includegraphics[height=8cm, width=8cm, angle=0]{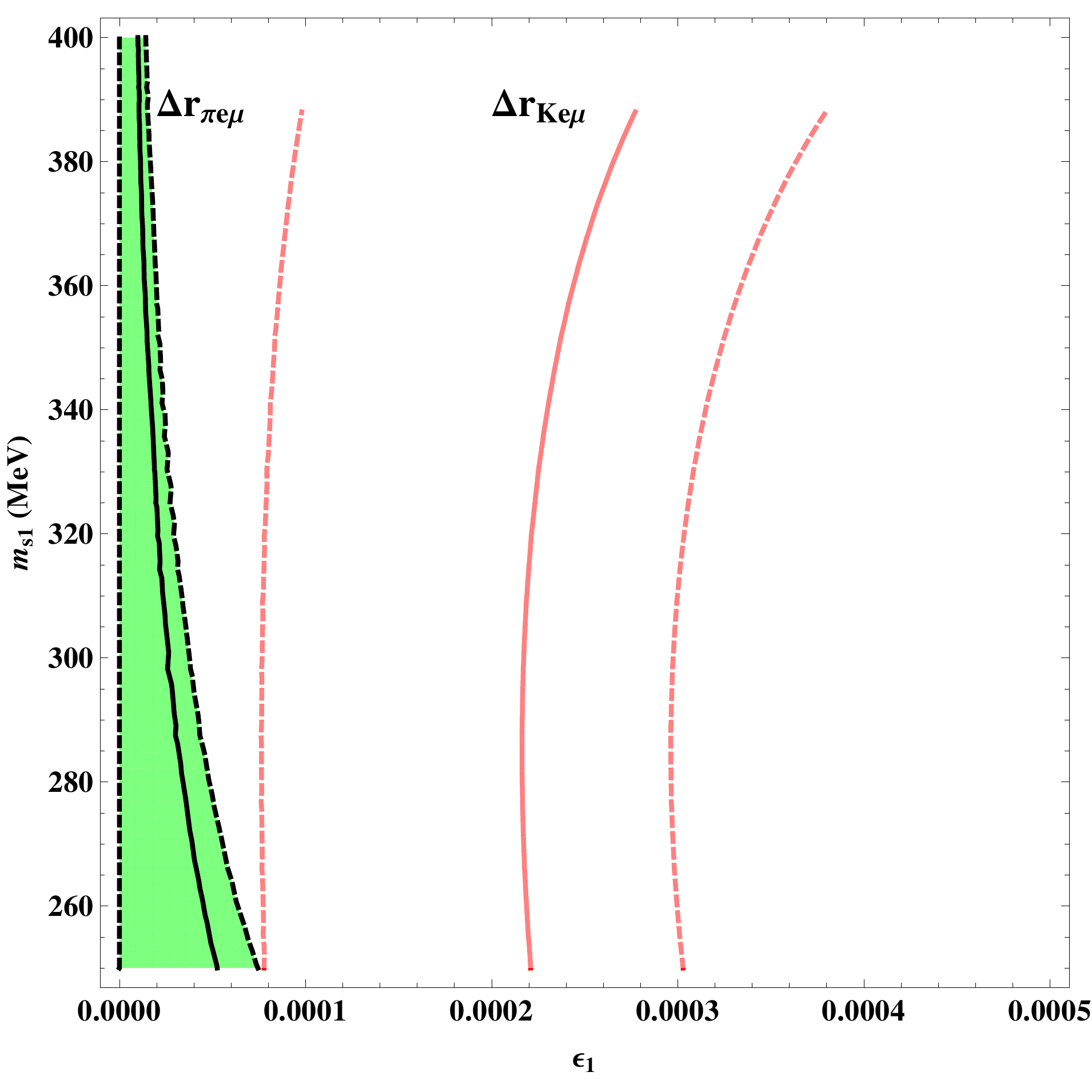}
\caption{Comparison of $\Delta r_{\pi e\mu}$ and $\Delta r_{Ke\mu}$ in the parameter spaces of mixing angles $\epsilon_1$ and the sterile neutrino mass $m_{s1}$.} \label{FigDrPiK}
\end{figure}

Besides, as shown in Fig. \ref{FigDrPiK}, the  3+1 scenario fails to provide a common parameter space of $\epsilon_1$ and $m_{s1}$ to saturate the experimentally measured  $\Delta r_{\pi e\mu}$ and $\Delta r_{K e\mu}$ simultaneously, namely there does not exist a solution within 1$\sigma$ tolerance.

\begin{figure}[t]
\centering
\includegraphics[height=12cm, width=12cm, angle=0]{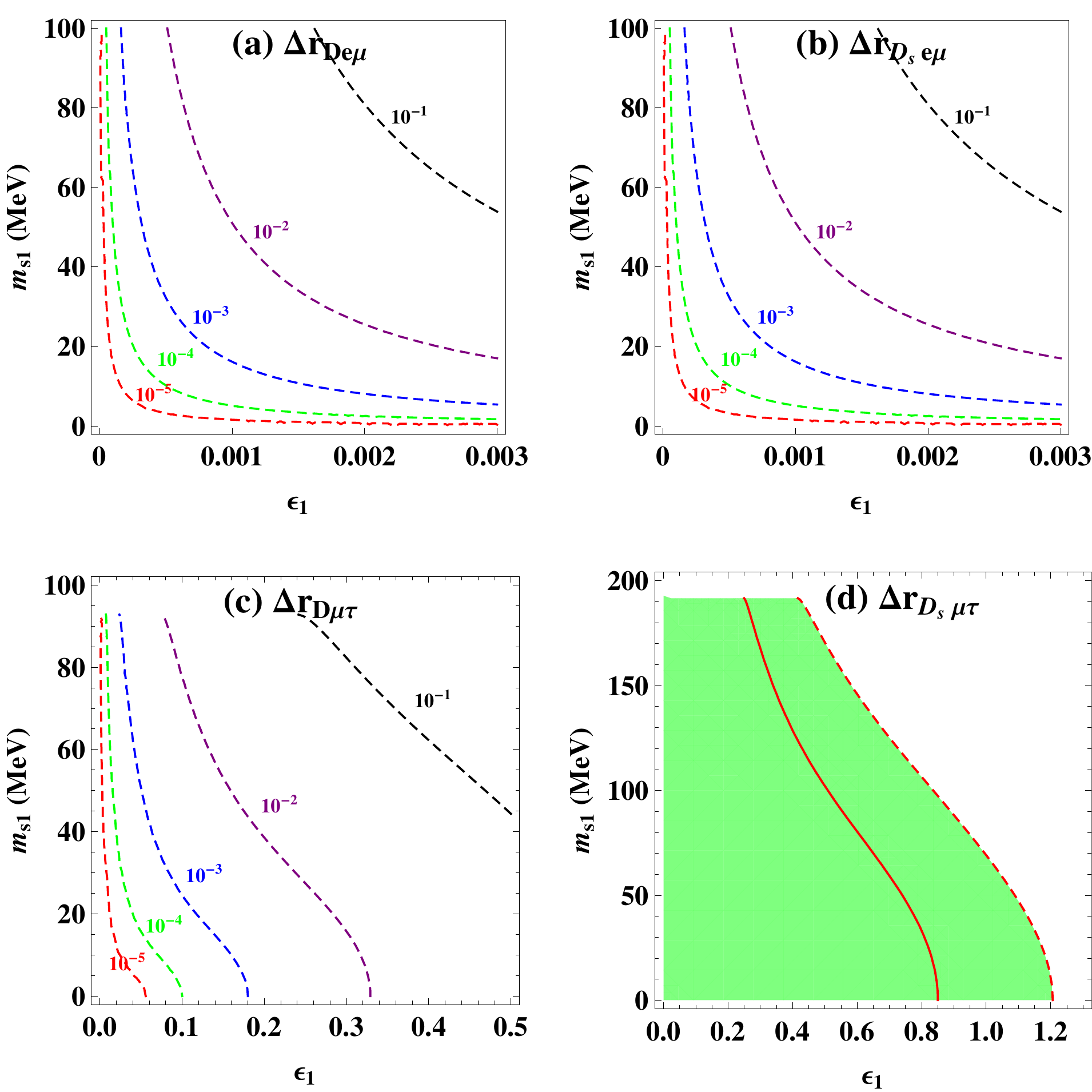}
\caption{$\Delta r_{D e\mu}$, $\Delta r_{D_s e\mu}$, $\Delta r_{D \mu\tau}$, $\Delta r_{D_s \mu\tau}$ vs $\epsilon_1$ and  $m_{s1}$.} \label{FigDrDDs}
\end{figure}

Since both $D_s$ to $\mu\nu$ and $\tau\nu$ have been experimentally measured, thus  $\Delta r_{D_s \mu\tau}$  is obtained. Lack of experimental data on the decay rates of leptonic decays of $D$, we cannot determine $\Delta r_{D \mu\tau}$ yet, for an illustration, we set the $\Delta r_{D\mu\tau}$ as $10^{-1}, 10^{-2}, 10^{-3}, 10^{-4}, 10^{-5}$ to get a sense about the dependence  of the $\Delta r$s on $\epsilon_1$ and $m_{s1}$. The results are shown in Fig. \ref{FigDrDDs}.

Form the figures of $\Delta r_{De\mu}$, $\Delta r_{D_se\mu}$ and $\Delta r_{D \mu\tau}$ in Fig. \ref{FigDrDDs}, it is obvious that non-zero $\Delta r$ demands non-vanishing $\epsilon_1$. Moreover, it is noted that for fixed $m_{s1}$ ($\epsilon_1$), the smaller $\Delta r$ is, the smaller $\epsilon_1$ ($m_{s1}$) would be.

In (d) of Fig. \ref{FigDrDDs}, the lower bound of $\Delta r_{D_s\mu\tau}=0.256_{-0.317}^{+0.430}$ does not appear, as our analyses indicate that $\Delta r_{D_s\mu\tau}$ cannot be negative in the 3+1 scenario. From this diagram, it is obvious that within $1-\sigma$ range of $\Delta r_{D_s \mu\tau}$, the particular values $\epsilon_1=0$ and $(m_{s1}=0)$ are not excluded and the vanishing sterile-active neutrino mixing signifies that the lepton flavor universality holds. Thus to make a decisive judgement  more accurate measurements are needed.

\begin{figure}[t]
\centering
\includegraphics[height=12cm, width=12cm, angle=0]{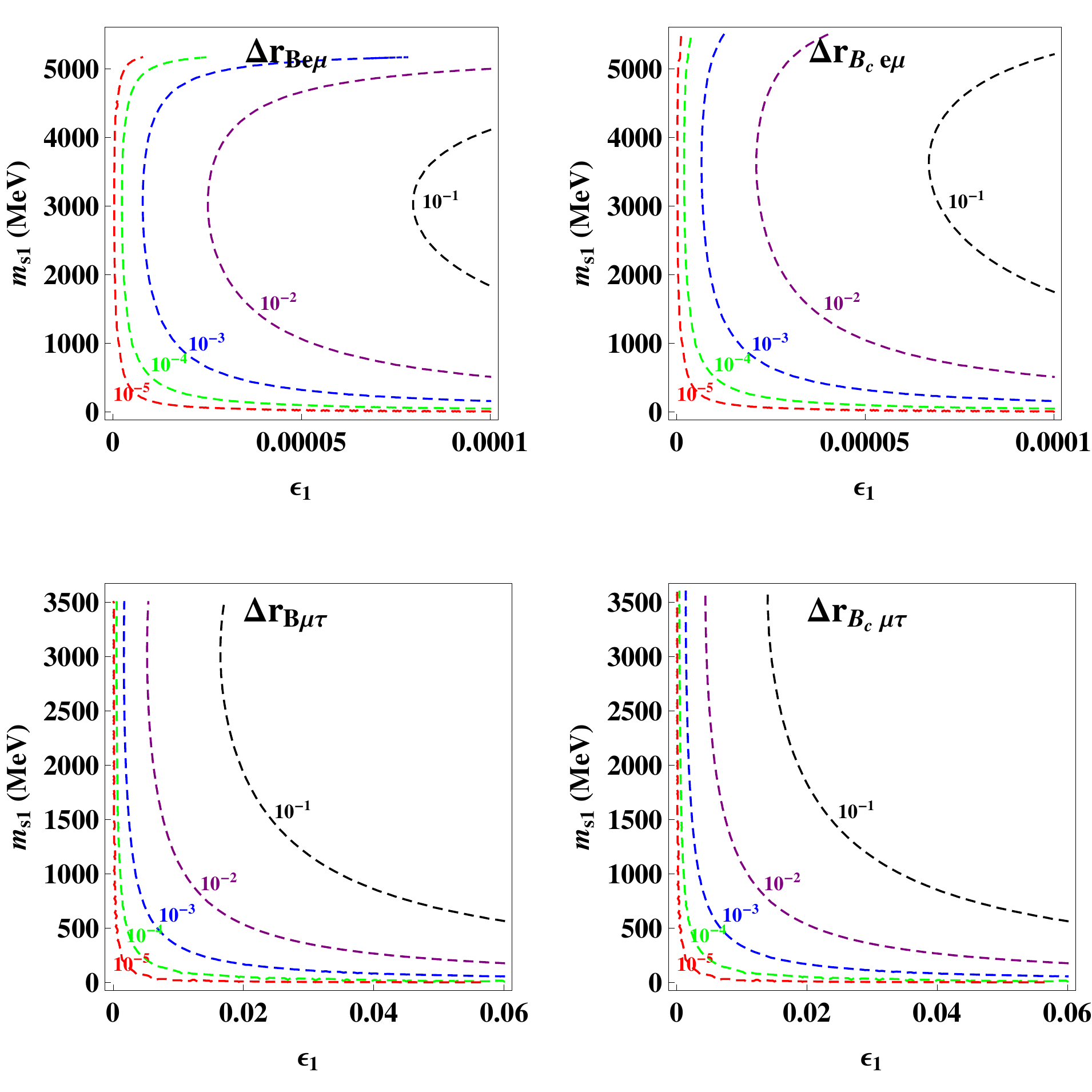}
\caption{$\Delta r_{B e\mu}$, $\Delta r_{B_c e\mu}$, $\Delta r_{B \mu\tau}$, $\Delta r_{B_c \mu\tau}$ in the parameter spaces of  $\epsilon_1$ and  $m_{s1}$ for different values.} \label{FigDrBBc}
\end{figure}

As discussing leptonic decays of $B$ and $B_c$ mesons, because of lack of experimental data, we take several values for $\Delta r$ to illustrate its dependence on the parameters as shown in Fig. \ref{FigDrBBc}.

\subsection{3+2 scenario}
In this subsection, we numerically analyze the lepton universality with two sterile neutrinos, i.e., the 3+2 scenario.

\begin{figure}
\centering
\includegraphics[height=10cm, width=10cm, angle=0]{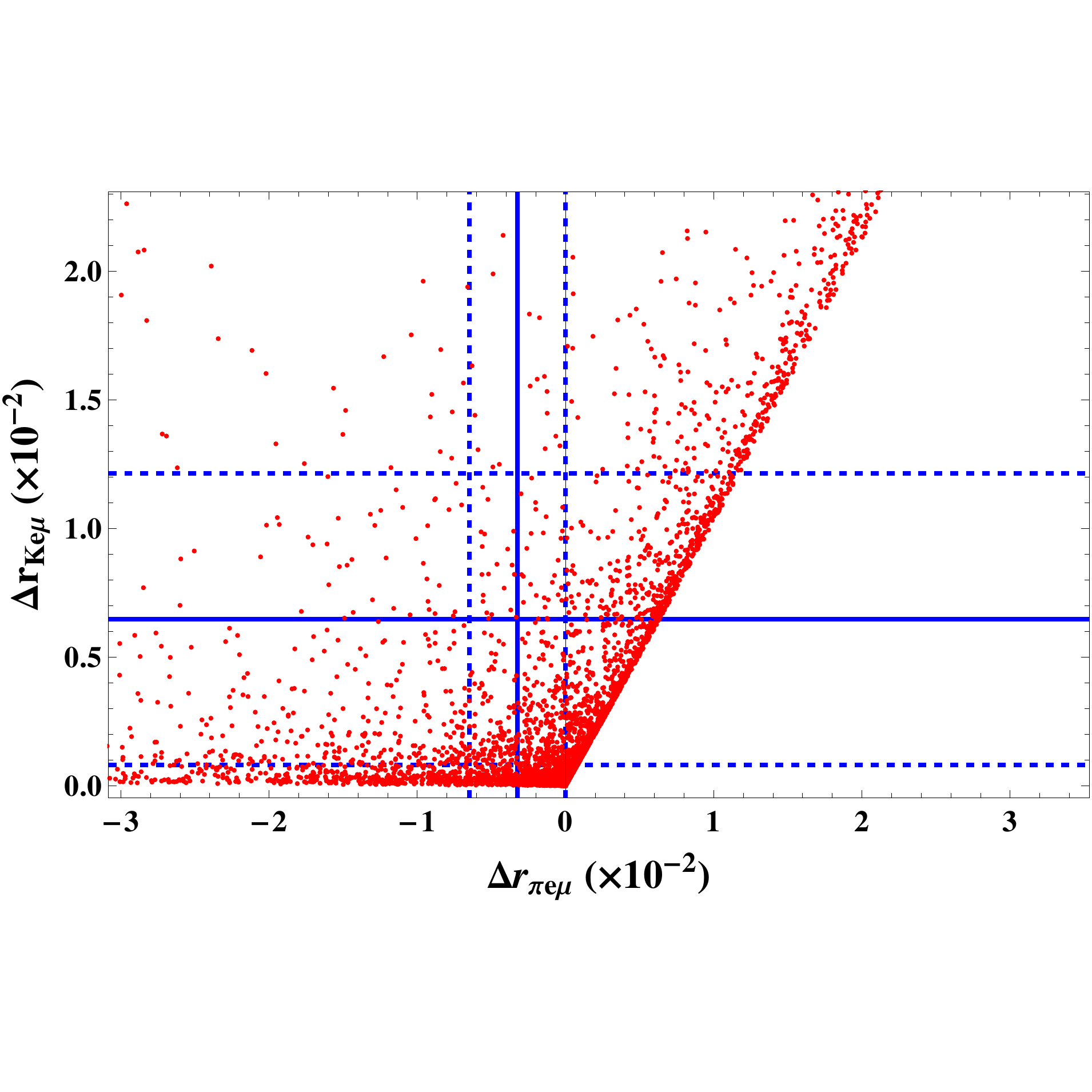}
\caption{The common solution for $\Delta r_{\pi e\mu}$ and $\Delta r_{K e\mu}$ in the 3+2 scenario.} \label{FigDrPiK2s}
\end{figure}

Even though the errors are still large, $\Delta r_{\pi (K) e\mu}$ have been set, thus we first present $\Delta r_{\pi e\mu}$ and $\Delta r_{K e\mu}$ in the same graph  Fig. \ref{FigDrPiK2s}, where the horizontal solid line corresponds to the central value and dashed lines enclose the $1-\sigma$ range of $\Delta r_{K e\mu}$ whereas the perpendicular ones are for $\Delta r_{\pi e\mu}$.

The cross region satisfies both $\Delta r_{\pi e\mu}$ and $\Delta r_{K e\mu}$. In our analyses, we let the mixing angles $\epsilon_1$ and $\epsilon_2$ vary within $(0, 3\times 10^{-3})$ and $(0, 5\times 10^{-5})$, respectively, while the sterile neutrino masses $m_{s1}\in (0, 140)$ MeV and $m_{s2}\in (0, 500)$ MeV. It is noted that for such parameter ranges, there exist solutions to accommodate both $\Delta r_{\pi e\mu}$ and $\Delta r_{Ke\mu}$. Concretely, the red points which correspond to the values  calculated within the parameter ranges fall in the common region of these two quantities. Existence of solutions satisfying both $\Delta r_{\pi e\mu}$ and $\Delta r_{Ke\mu}$ signifies success of 3+2 model to explain the observed lepton universality violation in $\pi$ and $K$ decays. Now we turn to $D_s$ decays, whose rates have been experimentally measured even though not sufficiently accurate yet, to see if we are able to determine $\Delta r_{D_s\mu\tau}$.

\begin{figure}
\centering
\includegraphics[height=10cm, width=10cm, angle=0]{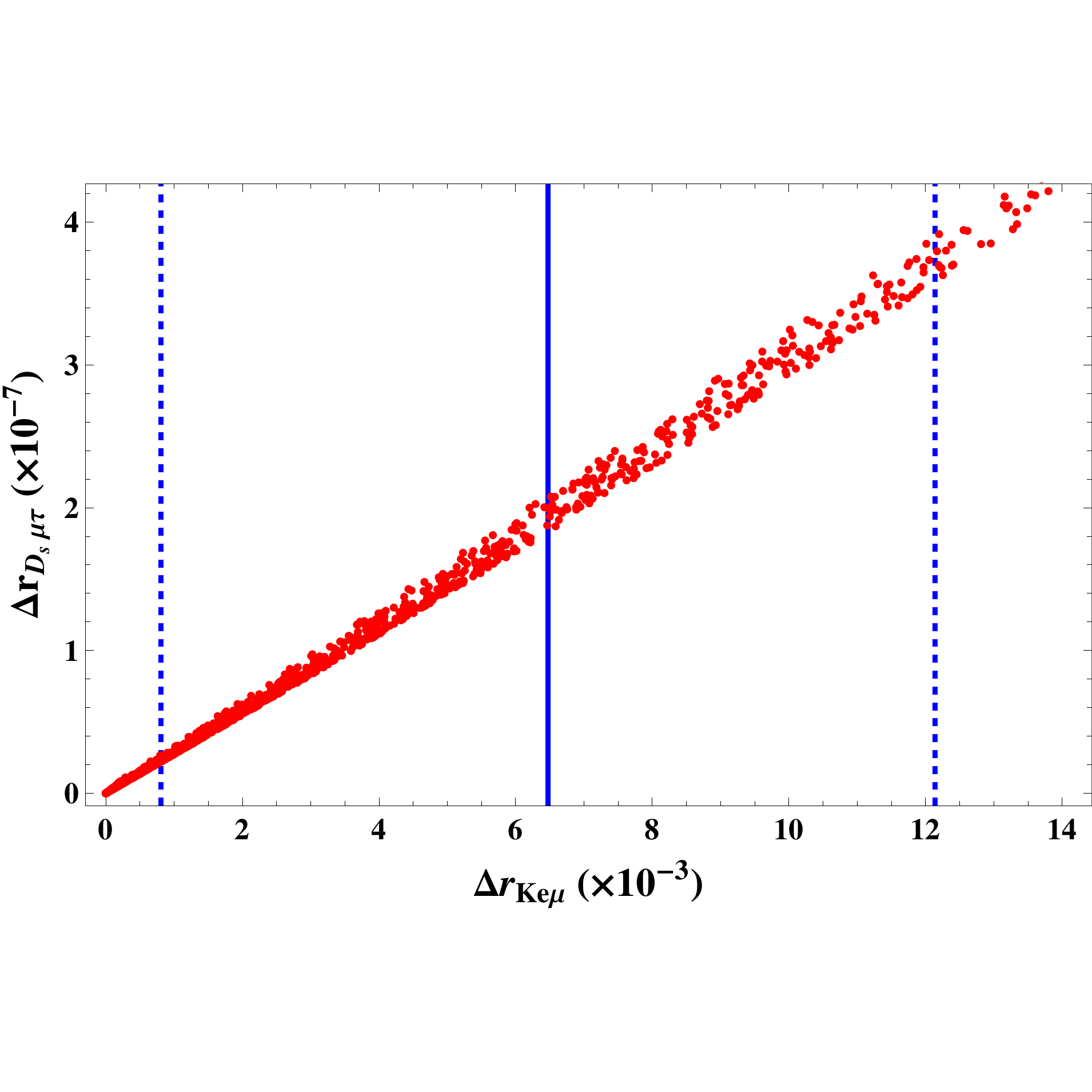}
\caption{The common solutions of $\Delta r_{K e\mu}$ and $\Delta r_{D_s \mu\tau}$ in 3+2 scenario.} \label{FigDrKDs2s}
\end{figure}

$\Delta r_{K e\mu}$ and $\Delta r_{D_s\mu\tau}$ are presented in Fig. \ref{FigDrKDs2s} where one notices that within the parameter ranges: $\epsilon_1\in (0, 6\times 10^{-4})$, $\epsilon_2\in (0, 5\times 10^{-5})$, $m_{s1}\in (1, 140)$ MeV, and $m_{s2}\in (0, 500)$ MeV, both $\Delta r_{K e\mu}$ and $\Delta r_{D_s\mu\tau}$ can be satisfied. Especially, in these parameter ranges  $\Delta r_{\pi e\mu}$ and $\Delta r_{K e\mu}$ are also satisfied.

\begin{figure}
\centering
\includegraphics[height=10cm, width=10cm, angle=0]{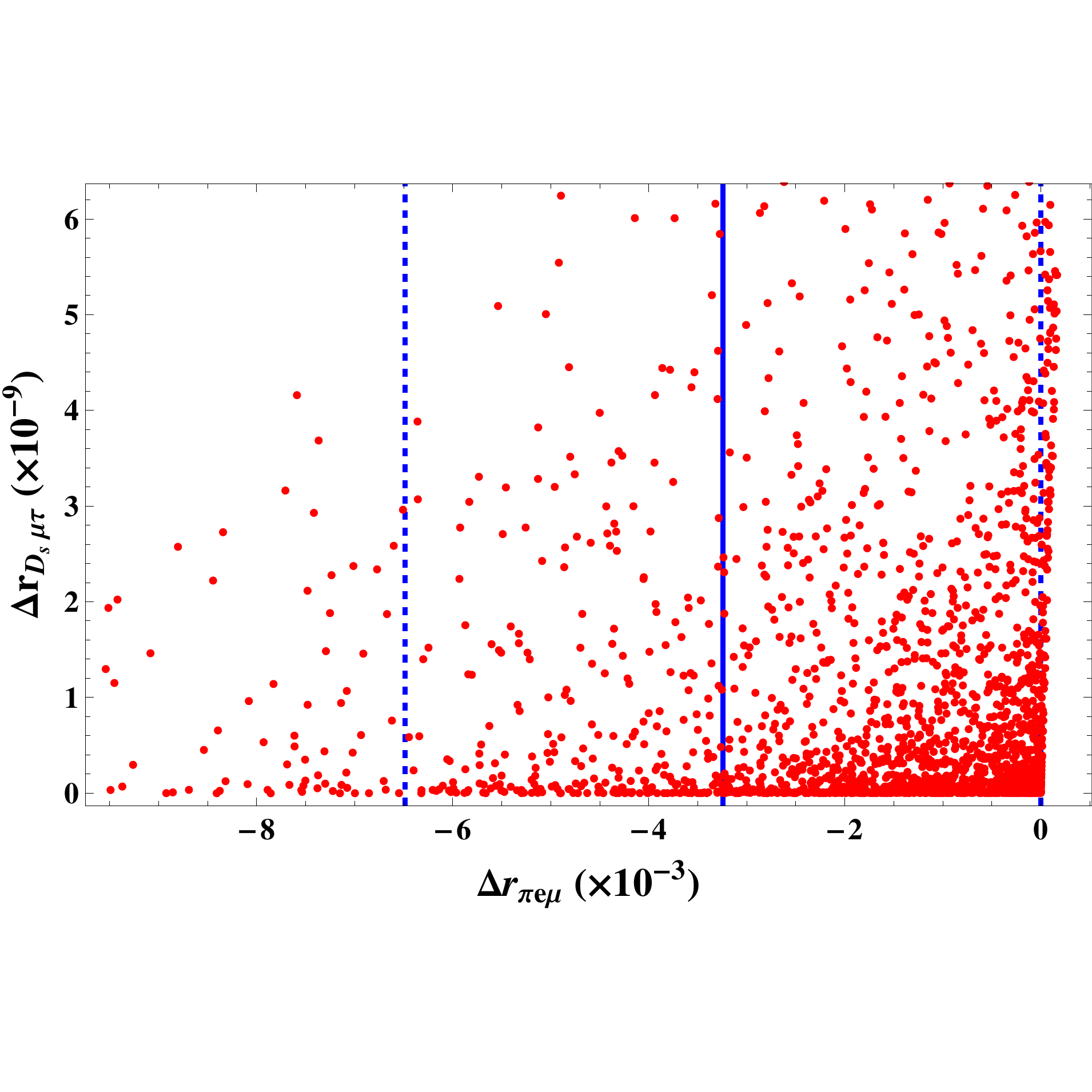}
\caption{The common region of $\Delta r_{\pi e\mu}$ and $\Delta r_{D_s \mu\tau}$ in 3+2 scenario.} \label{FigDrPiDs2s}
\end{figure}

The common region for $\Delta r_{\pi e\mu}$ and $\Delta r_{D_s \mu\tau}$ is shown in Fig. \ref{FigDrPiDs2s}.

\begin{figure}
\centering
\includegraphics[height=10cm, width=10cm, angle=0]{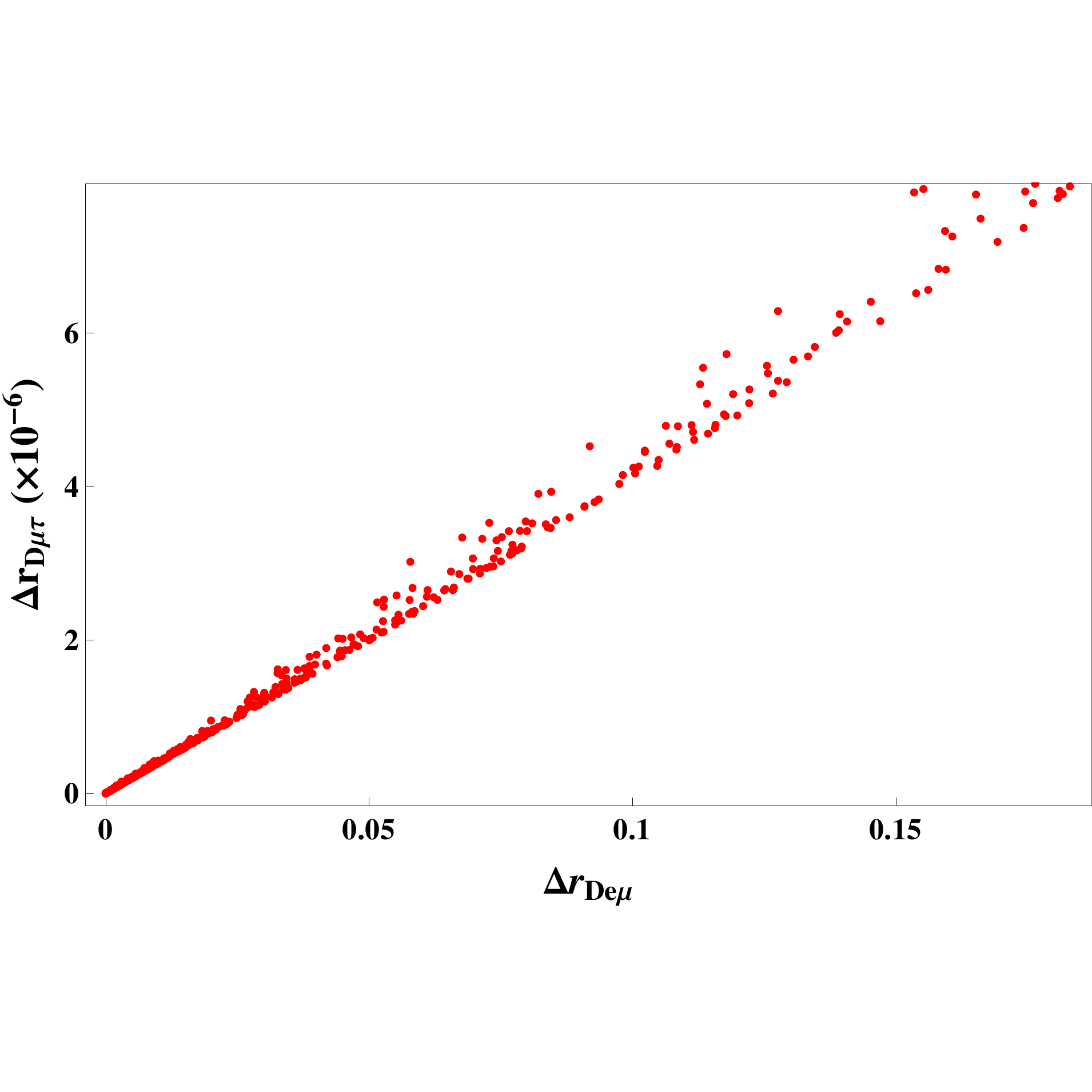}
\caption{The common solutions of $\Delta r_{D e\mu}$ and $\Delta r_{D \mu\tau}$ in 3+2 scenario.} \label{FigDrD2s}
\end{figure}

\begin{figure}
\centering
\includegraphics[height=10cm, width=10cm, angle=0]{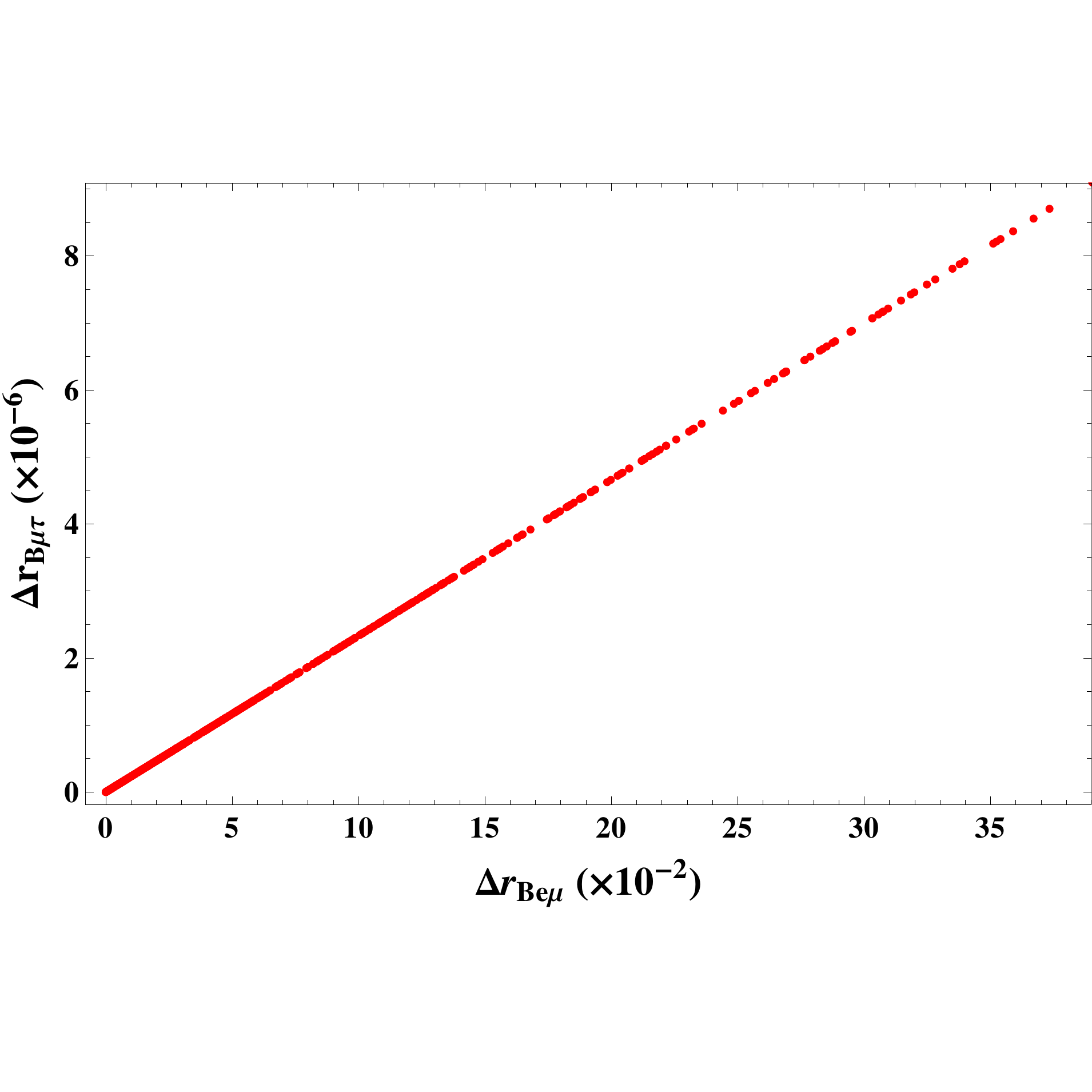}
\caption{The estimate of $\Delta r_{B e\mu}$ and $\Delta r_{B \mu\tau}$ in 3+2 scenario.} \label{FigDrB2s}
\end{figure}

The similar analyses have been carried out for $D$, $B$, and $B_c$ mesons. Due to shortage of data to calculate corresponding $\Delta r$s, we adopt the parameter ranges obtained by fitting $\Delta r_{\pi e\mu}$, $\Delta r_{K e\mu}$ and $\Delta r_{D_s \mu\tau}$, to the leptonic decays of $D, B$ and $B_c$ and investigate if there exist common region for $\Delta r_{Pe\mu}-\Delta r_{P\mu\tau}$ ($P=D, B$, or $B_c$). The results are shown in Fig. \ref{FigDrD2s} and Fig. \ref{FigDrB2s}.

\section{Testing LFU at BESIII}

Experimentally,  measurements on pure leptonic
decays of $D$ and $D_s$ mesons have been carried out by many collaborations:
via $e^+e^-$ annihilation at $Z^0$ mass pole
\cite{aleph, opal, L3}, at $\Upsilon(4S)$ mass
\cite{babar, belle}, and at
$\sqrt{s}=$ 3.773, 4.040 or 4.170 GeV \cite{cleo, bes},
respectively. To test violation of the lepton flavor universality, very high accuracy is necessary. But most of the the aforementioned experiments suffered from high background
contaminations, so do not meet the high accuracy demand. In this aspect, the electron-positron colliders prevail over others. Because  charmed mesons are
produced in pairs, one can accurately measure the pure leptonic
decays  based on the double tag method.

For example, the $e^+e^-$ annihilation experiment around the 3.773
GeV, where just above the $D\bar{D}$ production threshold, a charmed meson
and its anti-particle are produced in pair, i.e.
$\psi(3770)\to D\bar{D}$. If one fully identifies $\bar{D}$ in one event, which is called as a singly tagged $\bar D$ meson, there must exist a $D$ meson in the recoiling side against the tagged
$\bar{D}$ meson. And if one reconstructs the whole $D\bar D$ pair in the analysis, the event will be called as a double tag event. Thus, in an event which consists of a singly tagged $D^-$, the pure-leptonic modes can be selected from the final states of $D^+$ decays, and the absolute branching fractions would be well determined.

For the measurements around 4.040 or 4.170 GeV, situations are not much
different except $D\bar{D}$ being replaced by $D_S^+D_S^-$ or
$D_S^-D_S^{*+} + c.c.$.

At the BES III experiments, charmed mesons are collected at 3.773 and
4.040 GeV, respectively. Here we present a Monte Carlo (MC)
simulation at this two energy points to discuss the experimental
sensitivities of searching for pure-leptonic decays that can be
reached in the future.

The MC events are generated with the BES III offline Software System
\cite{boss}, where the particle trajectories are simulated with
the GEANT4 \cite{geant4} based package \cite{simbes} for
the BESIII detector \cite{bes3} at the BEPC-II collider.

The events used in this discussion are generated as
$e^+e^-\to\psi(3770)\to D\bar{D}$ and $e^+e^-\to\psi(4040)\to
D_S^+D_S^-$ at the c.m. energy $\sqrt{s}=$ 3.773 and 4.040 GeV,
respectively, where the $D\bar{D}$ and $D_S^+D_S^-$ mesons are set to decay into all possible final states with the branching fractions cited by PDG\cite{PDG}.

Totally $\sim1.23\times10^8$ $D\bar D$ and $\sim6.20\times10^6$
$D_S^+D_S^-$ events are generated at $\sqrt{s}=$ 3.773 and 4.040
GeV, respectively, corresponding to an integrated luminosity of
$\sim20$ fb$^{-1}$ $\psi(3770)$ and $\psi(4040)$ data assuming
$\sigma^{\text{obs}}_{D\bar{D}} = 6.14$ nb\cite{obs-dd} and
$\sigma^{\text{obs}}_{D_S^+D_S^-} = 0.31$ nb\cite{obs-dsds}, which
contains $\sim 7.2\times10^7$ $D^0\bar{D^0}$ pairs,
$\sim5.1\times10^7$ $D^+D^-$ pairs and $6.20\times10^6$ $D_S^+D_S^-$
pairs respectively. The BEPC-II collider is designed to work with an
instantaneous luminosity of $10^{33}$ cm$^{-2}$s$^{-1}$ around
$\psi(3770)$. As a conservative estimate, a data sample with an
integrated luminosity of about 20 fb$^{-1}$ can be collected during
more than 10 years' running.

The singly tagged $D^-$ and $D_S^-$ events are reconstructed in 9
hadronic decays of $D^-\to K^+\pi^-\pi^-$(50\%), $D^-\to
K_S^0\pi^-$(52\%), $D^-\to K_S^0 K^-$(48\%), $D^-\to
K^+K^-\pi^-$(40\%), $D^-\to K^+\pi^-\pi^-\pi^0$(28\%), $D^-\to
\pi^+\pi^-\pi^-$(56\%), $D^-\to K^0_S\pi^-\pi^0$(27\%), $D^-\to
K^+\pi^+\pi^-\pi^-\pi^-$(21\%), $D^-\to K^0_S\pi^-\pi^-\pi^+$(31\%)
and 9 hadronic decays of $D_S^-\to K_S^0K^-$(46\%), $D_S^-\to
K^+K^-\pi^-$(39\%), $D_S^-\to K^+K^-\pi^-\pi^0$(12\%), $D_S^-\to
K^0_SK^+\pi^-\pi^-$(24\%), $D_S^-\to \pi^-\pi^+\pi^-$(52\%),
$D_S^-\to \pi^-\eta, eta\to\gamma\gamma$(41\%), $D_S^-\to
\pi^-\eta^{\prime}, \eta^{\prime}\to\pi^+\pi^-\eta,
\eta\to\gamma\gamma$(21\%), $D_S^-\to \pi^-\eta^{\prime},
\eta^{\prime}\to\gamma\rho^0$(34\%), $D_S^-\to\rho^-\eta$(17\%)
constituting approximately 29\% of all $D^-$ decays and 30\% $D_S^-$
decays, respectively, where the numbers in brackets are
reconstruction efficiencies.

Tagged $D^-$ and $D_S^-$ events are selected by two kinematic
variables based on the principles of energy and momentum
conservations: (1) Difference in energy
\begin{eqnarray}
\Delta E\equiv E_{\text{f}}-E_{\text{b}},
\end{eqnarray}
where $E_{\text{f}}$ is the total energy of the
daughter particle from $D^-$ or $D_S^-$ in one event and $E_{\text{b}}$ is the $e^+/e^-$ beam energy for the experiment, is recorded to
describe the deviation from energy conservation caused by
experimental errors. (2) Beam-constrained mass
\begin{eqnarray}
M_{\text{BC}}\equiv\sqrt{E_{\text{b}}^2-(\Sigma_i \overrightarrow{p}_i)}
\end{eqnarray}
is calculated to reduce an uncertainty brought by experimental errors when measuring the momenta of the produced particles. By this definition, the energy $E_{\text{f}}$ in the expression of $$M^2_{\text{inv.}}\equiv E^2_{\rm f}-p^2_{\text{f}}$$ for the $\bar D$ invariant mass is replaced by $E_{\rm b}=E_{\rm c.m.}/2$, where $E_{\text{c.m.}}$ is the c.m. energy at which $D^+D^-$ pair is produced.

The total energy and momentum of all the daughter particles in $D^{\pm}$
and $D_S$ decays must satisfy the Energy Conservation (EC) principle, generally one needs to introduce a kinematic fit, including energy and momentum constraints and some relevant corrections, to reject those not satisfying EC, but being recorded due to an uncertainty of experimental measurement. This replacement of the real invariant mass by $M_{\text{BC}}$ partly plays the role.

Moreover, events are rejected if they fail to satisfy the selection
constraint $|\Delta E|<3\times\sigma_{\Delta E}$, which is tailored
for each individual decay mode, and $\sigma_{\Delta E}$ is the standard deviation of the $\Delta E$ distribution.

As the $D^{\pm}$ and $D_s$ events are correctly tagged, a peak in $M_{\rm BC}$ spectrum would emerge at the position of $D^-$ or $D_S^-$ mass. Finally, if there are more than one combinations in one tagged event, the one with the smallest $|\Delta E|$ is retained. After considering the detection efficiencies of each tag mode, $10837045\pm6122$ and $549811\pm1593$ tagged $D^-$ and $\bar D_S$ events are obtained based on data samples of about 20 fb$^{-1}$, respectively.

At the recoiling side against the tagged meson, the other charmed meson decays into a charged lepton and a neutrino. Since the neutrino does not electro-magnetically interact with detector matter, it cannot be recorded in the detector and contributes a missing energy, therefore,  a kinematic quantity is defined \begin{eqnarray}
M^2_{\rm miss}\equiv(E_{\rm
b}-E_{l^+})^2-(-\overrightarrow{p}_{D^-/D_S^-}- \overrightarrow{p}_{l^+})^2,
\end{eqnarray}
where $\overrightarrow{p}_{D^-/D_S^-}$ is the three-momentum of the fully reconstructed $D^-/D_S^-$, and $E_{l^+}$ ($\overrightarrow{p}_{l^+}$) is the energy (momentum) of the candidate lepton. The spectrum of $M^2_{\text{miss}}$ for the signal events should produce a peak near zero because the neutrino mass is very tiny.

To select the purely leptonic decay event of $D^+\to l^+\nu_l$, only one charged track identified as electron/muon/pion and no isolated photons are allowed at the recoiling side.

For the rare process of $D^+\to e^+\nu_{e}$, the signal number is determined to be 1 by counting the signal window of $M^2_{\rm miss}$. We set an upper limit on the number of signal events for $D^+\to e^+\nu_{e}$ to be 4.36 by using the Feldman-Cousins method \cite{feldman} in absence of background at 90\% confidence level. The upper limit on the branching fraction for $D^+\to e^+\nu_{e}$ is $\mathcal{B}(D^+\to e^+\nu_{e})<8.5\times10^{-7}$.

For the decay of $D^+\to \mu^+\nu_{\mu}$, the number of simulated signal events is obtained to be $N^{\rm obs}_{D^+\to\mu^+\nu_{\mu}}= 2611.1\pm55.0$ by fitting the $M^2_{\rm miss}$ distribution, and the efficiency for reconstructing $D^+\to\mu^+\nu_{\mu}$ against the tag side is estimated to be $N^{\rm obs}_{D^+\to\mu^+\nu_{\mu}}= (63.82\pm0.15)\%$. Therefore, the branching fraction for $D^+\to\mu^+\nu_{\mu}$ is calculated to be $\mathcal{B}(D^+\to\mu^+\nu_{\mu}) = (3.74\pm0.08(\text{stat.}))\times10^{-4}$.

For $D^+\to\tau^+\nu_{\tau}$, with $\tau^+\to\pi^+\overline{\nu}_{\tau}$, the missing mass squared $M^2_{\rm miss}$ for the candidate events is calculated as
\begin{eqnarray}
M^2_{\text{miss}}\equiv(E_{\text{b}}-E_{\pi^+})^2 -
(-\overrightarrow{p}_{D^-}-\overrightarrow{p}_{\pi^+})^2.
\end{eqnarray}
Since there are two missing neutrinos, the $M^2_{\rm miss}$ does not have a narrow peak as in the decay of $D^+\to \mu^+\nu_{\mu}$. To suppress the background from $D^+\to K^0_L\pi^+$ by missing $K^0_L$ and $D^+\to\mu^+\nu_{\mu}$ by $\mu/\pi$ mis-identification, the $M^2_{\text{miss}}$ is studied in two cases defined by $E^{\rm EMC}_{\pi}>0.3$ and $E^{\text{EMC}}_{\pi}<0.3$ GeV, resulting the number of signal events for $D^+\to\tau^+\nu_{\tau}$ to be $312.1\pm28.2$ and $242.9\pm26.1$, respectively. With these signal events, inputting the detection efficiency of $(17.16\pm0.14)\%$ and $(13.33\pm0.12)\%$, the branching fraction is determined to be $\mathcal{B}(D^+\to\tau^+\nu_{\tau}) = (1.68\pm0.12(\text{stat.})) \times10^{-3}$, which is averaged by $(1.68\pm0.16(\text{stat.}))\times10^{-3}$ and $(1.68\pm0.18(\text{stat.}))\times10^{-3}$ for the two cases.

Following the similar analysis, the number of signal events for $D_S^+\to l^+\nu_l$ is determined to be $N^{\rm obs}_{D_S^+\to e^+\nu_{e}}= 1$,  $N^{\text{obs}}_{D_S^+\to \mu^+\nu_{\mu}}= 2335.5\pm55.1$ and $N^{\text{obs}}_{D_S^+\to \tau^+\nu_{\tau}}= 11484.1\pm110.2$, with the corresponding detection efficiencies to be $\epsilon^{\text{obs}}_{D_S^+\to e^+\nu_{e}}= (46.94\pm0.16)\%$, $\epsilon^{\text{obs}}_{D_S^+\to \mu^+\nu_{\mu}}= (70.41\pm0.14)\%$ and $\epsilon^{\text{obs}}_{D_S^+\to \tau^+\nu_{\tau}}= (39.41\pm0.16)\%$, respectively. Inserting the numbers of events and upper limit at 90\% C.L. for $D_S^+\to e^+\nu_{e}$, the branching fractions are calculated to be $\mathcal{B}(D_S^+\to e^+\nu_{e})<1.69\times10^{-5}$, $\mathcal{B}(D_S^+\to \mu^+\nu_{\mu}) = (6.03\pm0.14(\text{stat.}))\times10^{-3}$, and $\mathcal{B}(D_S^+\to \tau^+\nu_{\tau}) = (5.30\pm0.05(\text{stat.}))\%$.

Based on these simulation results, the ratio of decay rates to different leptons can be obtained to test the lepton universality. With 20 fb$^{-1}$ BES III data samples at c.m. energy of 3.773 and 4.040 GeV, the ratios are
\begin{eqnarray}
R_{De\mu}=\frac{\Gamma(D^+\to e+\nu_{e})}{\Gamma(D^+\to\mu+\nu_{\mu})}<2.26\times10^{-3},
\end{eqnarray}
and
\begin{eqnarray}
R_{D\mu\tau}=\frac{\Gamma(D^+\to\mu+\nu_{\mu})} {\Gamma(D^+\to\tau+\nu_{\tau})}=0.223\pm 0.017(\text{stat.})
\end{eqnarray}
for charm meson, and
\begin{eqnarray}
&&R_{D_Se\mu}=\frac{\Gamma(D_S^+\to e+\nu_{e})}{\Gamma(D_S^+\to\mu+\nu_{\mu})}<2.80\times10^{-3},\\
&&R_{D_S\mu\tau}=\frac{\Gamma(D_S^+\to\mu+\nu_{\mu})} {\Gamma(D_S^+\to\tau+\nu_{\tau})}=0.114\pm0.003(\text{stat.})
\end{eqnarray}
for the strange-charmed meson. The experimental sensitivities for the above measurement with 10 years' (20 fb$^{-1}$) data accumulation are $\Delta_{R^{\tau/\mu}_D}\sim7.35\%$ and $\Delta_{R^{\tau/\mu}_{D_S}}\sim2.50\%$, however, the size of these huge data samples cannot present a solid estimate for the electron decays. The theoretical expectation for electron mode is at $10^{-5}$ level, to challenge this limit, there should be a desperate running time for BES III experiment. Fortunately, it will not be a problem if one can build a $\tau$-charm factory with an increasing of the luminosity of about 100 times.

The luminosity of BEPC II is not high enough, so that to fulfill the job, one needs at least 10 years machine running. However as suggested, the planned charm-tau factory will greatly enhance the luminosity and with the new facility, we expect that in a few months a sufficiently large database could be collected and then one may have required accuracy to testify the lepton universality.

\section{Discussions and conclusions}
In this work we study the lepton universality in the  3+1 and 3+2 scenarios. The analyses indicate that by adding one sterile neutrino to the SM, i.e., the 3+1 scenario, the $\Delta r$s reflecting the differences between experimental data and SM predictions in the leptonic decays of various pseudoscalar mesons cannot be accommodated simultaneously for various mesons. Therefore, the 3+1 scenario is attractive for its simplicity, but does not meet the data. Whereas the 3+2 scenario has less tension in accordance with the data. However, although there still exists difficulty in choosing proper a mass range for the second sterile neutrino, the current experimental data on $\Delta r_{\pi e\mu}$, $\Delta r_{Ke\mu}$, and $\Delta r_{D_s\mu\tau}$ can be well accommodated. This result motivates one to incline to involving more sterile neutrinos, i.e., the 3+3 model. With the parameter ranges selected by fitting the data for $\pi$, $K$ and $D_s$, the predicted values of $\Delta r$s for $D$, $B$, and $B_c$ are obtained which will be tested by the future experiment such as the Z-factory and LHCb.

Usually it is believed that the $\mu-\tau$ symmetry \cite{mutau} holds at high-energy scales, but breaks during the evolution to low-energy, so that the $3\times 3$ PMNS mixing matrix for active neutrinos deviates from the original symmetric textures.  The violation of lepton universality, especially that for $\mu-\tau$ universality, might be a low-energy behavior as the universality precisely holds at high-energy scale, say, the GUT  or the see-saw energy scales. With these assumptions, it motivates one to relate the breaking of the $\mu-\tau$ symmetry to the violation of the $\mu-\tau$ universality. The idea is that the symmetry breaking leading to real PMNS matrix and the LFU violation may originate from same source and caused by the smae mechanism. Thus both of them serve as the low-energy manifestations of the symmetry beaking.

In order to obtain a negative value for $\Delta r$, $R_{P\alpha\beta}^s$ should be smaller than the value predicted by the standard model  $R_{P\alpha\beta}^{\text{SM}}$. Then one can obtain
\begin{eqnarray}
R_{P\alpha\beta}^{\text{SM}} < {{\sum_{j=1}^{N-3}|U_{\beta, j+3}|^2(m_\beta^2+m_{sj}^2)(m_P^2-m_\beta^2-m_{sj}^2)\lambda(m_P^2, m_\beta^2, m_{sj}^2)}\over{\sum_{i=1}^{N-3}|U_{\alpha, i+3}|^2(m_\alpha^2+m_{si}^2)(m_P^2-m_\alpha^2-m_{si}^2)\lambda(m_P^2, m_\alpha^2, m_{si}^2)}},
\end{eqnarray}
i.e., the ratio of the contributions from sterile neutrino should be larger than the SM prediction $R_{P\alpha\beta}^{\text{SM}}$.

Obviously, ignoring high order QED radiative corrections, only the mass of the concerned pseudoscalar meson enters the game, but not the identities of its constituents, thus we can relate $\Delta r$s of various mesons to each others. This conclusion is viable for checking the  scenarios discussed in the introduction.


Our numerical analyses indicate that checking violation of LFU presents a rigorous requirement for our colliders, therefore a high-luminosity charm-tau factory and/or B-factory are necessary to draw solid conclusion.

\begin{acknowledgments}
This work is supported by the National Natural Science Foundation of China under the contract No.11075079, 11135009.
\end{acknowledgments}

\end{document}